# Fabrication of 3D Packaging TSV using DRIE

M.Puech, JM Thevenoud , JM.Gruffat, N. Launay, N. Arnal, P. Godinat,
**Alcatel Micro Machining Systems**
**Annecy, France**

## ABSTRACT

3D stacking of die with TSV (through Silicon Via) connection as well as wafer level packaging of CMOS image sensors (CIS) are becoming very hot topics. While TSV of CIS is definitively a back-end technique, 3D stacking of die through TSV can be done with different strategies: from the via first approach, a front-end process, to the via last approach, a back-end process. Each of these different ways of elaborating the vias has its advantages and drawbacks in terms of electrical performances, refilling materials and cost. They have in common the need to etch the vias. In this paper, we will review the DRIE performances for the definition of the different via shapes, depths, and sizes. A very low temperature (< 150 °C) PECVD process has also been characterized and will be presented for the high demanding packaging of CIS.

## INTRODUCTION

The ever storage capacity demanding products as MP3, mobile phones, digital still camera,.. are asking for smaller size packages and higher memory density.In response to the performance and density requirements, the semiconductor industry has definitively moved from 2D to 3D style packages with shorter electrical connections. Through silicon vias (TSV) and related techniques are enabling technologies clearly identified.

The fabrication of TSV includes three major steps: silicon via etching, via insulation and via metallization. Depending upon the positioning of the TSV sequence within the overall wafer manufacturing, there are many different process flows. In addition to the "Via First" approach, developed within the Emc3d consortium [1], we review the other "Via First before FEOL" and "Via Last after BEOL" alternative solutions.

## 1. DRIE PROCESS

Developed more than 10 years ago for the MEMS emerging products, the DRIE tools are based on ICP source because of their high degree of gas dissociation, Fig. 1.

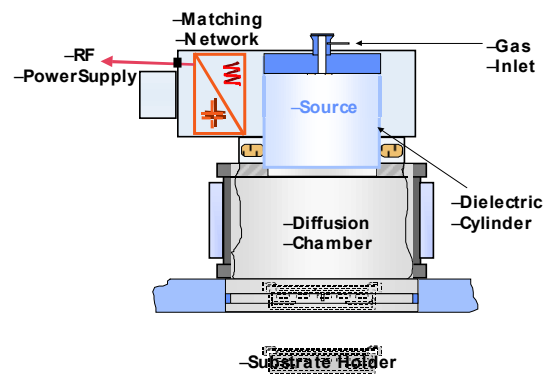

Fig. 1: AMMS de-coupled ICP process chamber

The most common used DRIE process is known as the "Bosch" process [2]. This process alternates short step of $SF_6$ plasma for the fast but isotropic removal of silicon, with short $C_4F_8$ plasma deposition step for the sidewalls protection; the polymer layer will be removed at the bottom of the feature during the first part of the next etching step with SF6. Thanks to the "F" based chemistry, this process has the potential to deliver very high and selective etching rate.

## 2. "VIA FIRST prior to FEOL" TSV

The "via first prior to FEOL" strategy is the strategy where the vias are done on the blank Si wafer prior to any CMOS process (Fig. 2). Due to the step position in the wafer fabrication process, it can be realized by the CMOS manufacturer, or even the wafer supplier, but not by the packaging company.

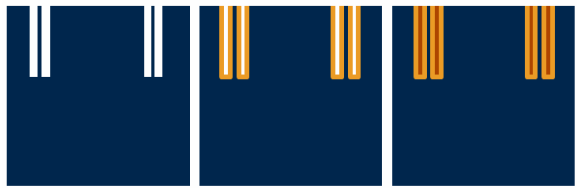

Fig. 2 : "Via first prior to FEOL"
From left to right: D.R.I.E., Oxidation, Poly-Si refilling

Due to all the further CMOS manufacturing steps, the refilling material has to be resistant to all the thermal processes, usually at higher than 1000 degrees. That's why the mostly used refilling material is poly-silicon. As the poly-silicon refilling process can be achieved only in narrow features, the etched features have a width lower than 5µm. One advantage of such process is that it doesn't require any seed layer, and the isolation layer can be easily done using a traditional oxidation process.

As the wafer will be thinned down at the latest stage to a thickness around 150µm, the less than 5µm wide features have to be etched down to at least 150µm, that means a more than 30 aspect ratio (Fig. 3). The Aspect Ratio (A.R.) of a given feature, like a trench, is defined as the ratio between the depth of the trench and the width.

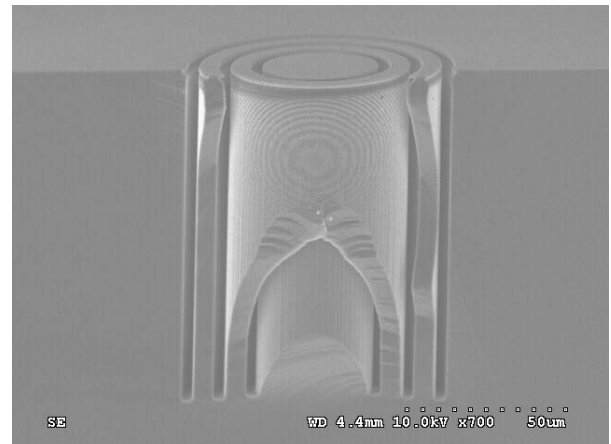

Fig. 3: Typical "Via first prior to FEOL" etch
Courtesy: ST Microelectronics

The use of un-pulsed processes, such as a mixture of SF6 and O2, is not possible to reach such high aspect ratio, and the selectivity is also too low. A "Bosch" process, or even better, a "S.H.A.R.P." (Super High Aspect Ratio Process) process [3] can be used to achieve the best etch performances. This Alcatel patented process allows an increase of the aspect ratio in comparison with the "basic" Bosch process". When applying the innovative "S.H.A.R.P." process to deep etching of very small features, it was demonstrated that an Aspect Ratio as high as 110 was achievable as shown in Fig. 4.

The D.R.I.E. is the only technology available that can etch features of 5µm wide to a depth higher than 100µm in Silicon, and Alcatel Micro Machining Systems is the only company that can provide the S.H.A.R.P. process.

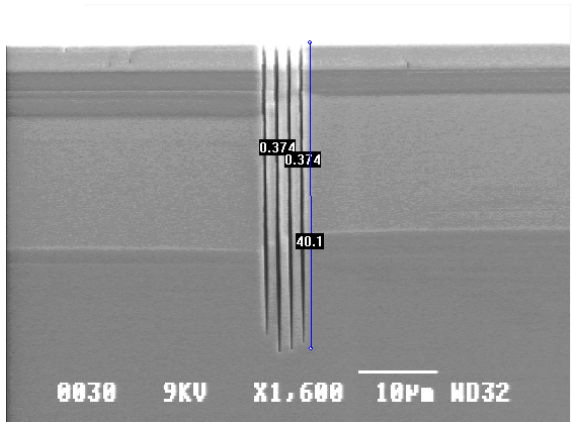

Fig. 4: *The "SHARP" Process for submicron features. (Courtesy:ESIEE)*

For a manufacturing point of view, the "via prior to FEOL" strategy has the big advantage to be easily integrated to the actual fabrication process, as there is only the D.R.I.E. equipment to be added to the current fabrication. All the other steps are achievable using the current tools used in a semiconductor fab. And so, the integration is very easy and not at all costly.

## 3. "VIA FIRST after BEOL":

In this approach, the TSV will be realized once the CMOS device will be completed and before the grinding process for wafer thinning (Fig. 5).

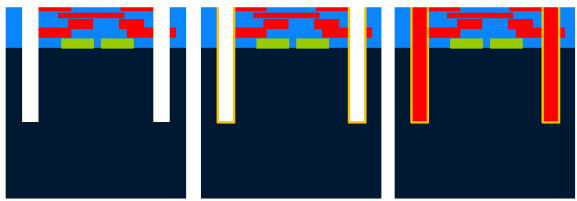

Fig. 5: "Via first after BEOL"
From left to right: D.R.I.E., Oxidation, Copper refilling

The advantage of this strategy is that the CMOS structure is already finished and passivated. At this stage, since the wafer will no longer be exposed to high temperature cycles, the use of via copper refilling material with much better electrical and thermal properties compared to poly-silicon is allowed.

Typical features size for those "Vias first after BEOL" are diameters ranging from 10 to 40µm etched at depth of 70-120µm. After etching and before refilling the via with copper, a dielectric layer and later a barrier and a seed layer have to be applied using CVD and PVD techniques (Fig.6).

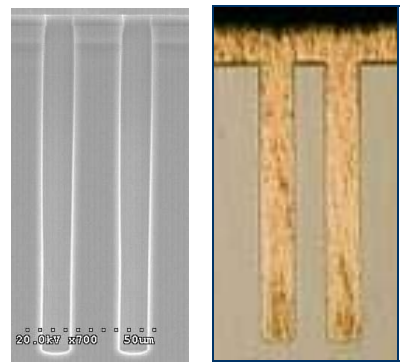

Fig. 6: Left "Via first after BEOL" after D.R.I.E., Right: "Via first after BEOL" after refilling step.

For a better coverage of the metal layer, a typical 83° to 85° tapered profile is usually required. As this step is quite costly and time consuming, an alternative "Open Mouth" profile has been successfully developed.

This new approach consists in making a facet at the top part of the hole, enlarging the entrance of the via hole. Starting with the same mask design previously used, the proposed process consists in a first step during which a highly tapered profile is applied through the use of mixed $SF_6 + C_4F_8$ plasma chemistry and then to apply a pure anisotropic process as shown on the Fig. 7. We evaluate this method targeting a facet angle of 70° on the first 30 µm of depth.

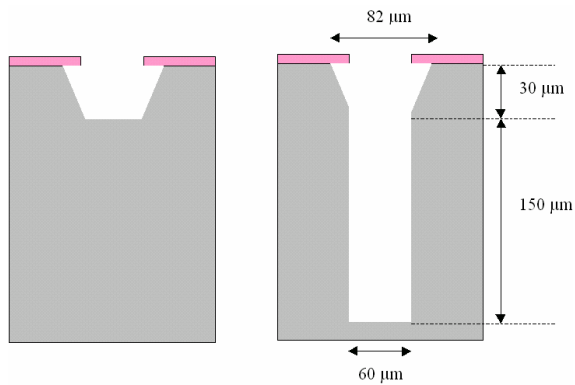

Fig. 7 : "Ideal" profile (top), Tapered etching step (left bottom), followed by vertical etching step (right bottom).

When the desired depth is achieved, the process recipe is switched to an anisotropic process. Because the anisotropic process is ion enhanced, the transition between tapered and vertical profile occurs exactly underneath the mask opening, keeping a good CD control. Fig. 8, shows the result of such a combined process: the average etch rate is still high at more than 12 μm/min with a very good uniformity (+/- 2.5%), the profile angle is 70° at the top 30 μm then 88,8° down to 180 μm depth.

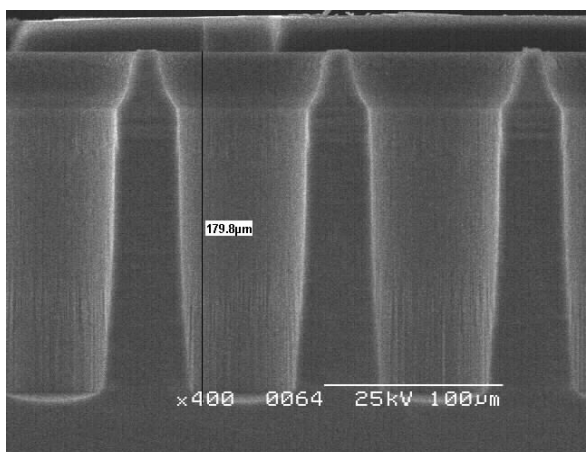

Fig. 8: 60x180 μm Via Hole; Profile angle:70° at top 30 μm then 88.8° down to 180 μm.

Because of the further thinning operation, the etched depth is limited to less than 100μm resulting in relatively low aspect ratios not exceeding 5 to 10, which are very favorable to high etching rate regimes required for high throughput manufacturing.

The fluorine radicals produced in the ICP plasma are transported within the gas phase to the silicon surface where they react with the silicon to produce volatile $SiF_4$ molecules evacuated by the pumping system according to:

$$Si(s) + 4F(g) \rightarrow SiF_4(g) + \Delta G_0 \text{ (eq.1)}$$

A given plasma parameters setting, as flow rate, pressure, ICP power.., will produce a given "F" partial pressure available for the removal of silicon. As the Si etch rate is directly proportional to the fluorine radicals reaching the substrate (Fig. 9) and thanks to its 20 years knowledge in plasma technology, Alcatel Micro Machining Systems has developed the hardware as well as specific processes to produce higher quantity of fluorine radicals.

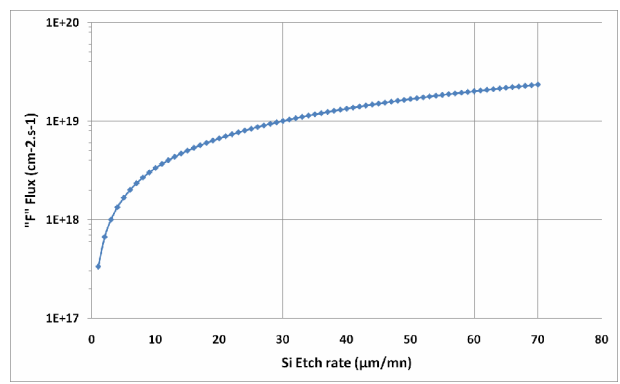

Fig. 9: Fluor radical flux rates versus Si etch rate

A high Si etch rate step combined with the most efficient passivation or deposition step, has allowed us to achieve anisotropic profiles, with very high etch rates of 50μm/min. or more.
As the etch rate increases, the capability to control the wafer temperature becomes crucial to obtain a uniform etch across the wafer whatever the wafer diameter.

In order to evacuate all this heat and achieve the best uniformity of temperature, Alcatel has designed a "P" type electrostatic chuck

(ESC) with an innovative design, allowing to adjust the temperature regarding the wafer patterns. For a given process, the "P" type ESC, gives a uniform temperature of ±0.15 degree across the wafer, with an excellent thermal conductivity.

## 4. "VIA LAST after BEOL":

Similarly to the wafer level packaging of the CMOS Image Sensors (CIS), in the "Via last" approach, the device wafer is grinded and thinned to its final thickness prior to the Via formation (Fig. 10). For that operation, the device wafer is temporary bonded onto a wafer carrier, usually made of glass. It will remain attached to this carrier until the stacking of the wafer onto another wafer, or until the transfer of the wafer onto a dicing frame, depending upon the attachment technique: wafer to wafer or die to wafer.

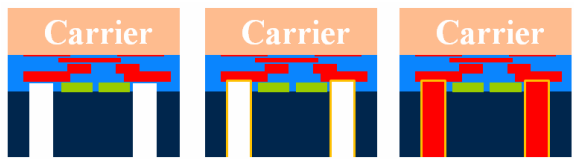

Fig. 10: "Via last after BEOL"
From left to right: D.R.I.E., Oxidation, Copper refilling. A temporary wafer carrier is used from the via formation until the complete refilling of the via.

Because the wafer carrier is little bit larger than the silicon wafer for a robust wafer, the overall handling system has to accept wafer substrates with diameter 2 mm larger than the standard wafers. Another major difference is on the capability of the wafer chuck to be able to electro statically clamp Silicon on Glass (SOG) substrates.

From a process perspective, the fact that the wafer is temporary bonded with adhesive onto a wafer carrier, implies that all the further process steps have to be limited at temperatures between 180 °C to 200 °C.

The DRIE "Via last" etching is very similar to the "Via First", after BEOL approach in term of via dimensions and profiles. However, we have to keep in mind that the wafer temperature limitation, discussed above, will be a major limiting factor in the high etching rate regime. In order to minimize the impact of the glass carrier onto the wafer surface temperature, we developed specific low thermal resistance ESC's.

### DRIE of high aspect ratio oxide:

The major difference of the "Via Last" approach compared to the "Via First" is that in this technique, the backside via metallization has to come in contact with the metal pad contact. Therefore, after the DRIE of the via, the stacked insulating oxide layers have to be dry etch removed down to the metal pad; then a conformal dielectric layer has to be deposited (at low temperature) and finally, this dielectric layer has to be cleared from the bottom via such to allow the deposition of the barrier and seed layers to be in contact with the metal pad. If the via profile is almost vertical, the two steps of dielectric etching can be done quite easily without any additional photolithography steps. If the via profile is quite tapered (for and easy PVD deposition of the barrier and seed layers), then the etching of the dielectric layers will necessitate a specific photolithography step to remove the dielectric layers from the bottom of the Via while keeping a thick sidewall layer for a good electrical isolation. In the packaging of the CIS, we already demonstrated that is was possible to remove the insulating layers underneath the metal pads using our high density plasma dielectric etcher. However in the packaging of CIS application, the via diameter is much larger than the one targeted for the die stacking (70μm to 80μm) and the taper angle is much pronounced ( 65° to 75°) allowing for easier photo-resist coating and developing. In order to demonstrate that we could extend our technology to higher aspect ratio features as the ones involved in the die stacking, we evaluated the etching of a buried oxide layer just after the DRIE of Si.

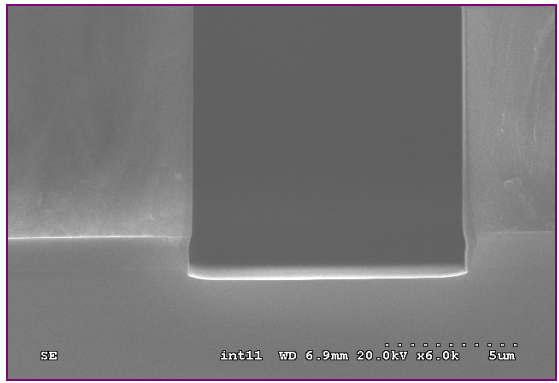

Fig. 11: Etching profile of a buried oxide layer underneath a Silicon layer.

The buried oxide layer was removed at 500 nm/mn with a 10:1 selectivity compared to the P.R. mask used to define the Via (Fig. 11).

**Low Temperature PECVD of dielectrics:**

As for the other process steps, the deposition of a conformal insulating layer has to be done while keeping the substrate temperature at around +180°C. Taking benefit of our experience in the deposition of silicon nitride passivation layers for the manufacturing of III-V compounds, we developed specific LT-PECVD of oxide layers adapted at larger wafer diameter.
Based on the AMMS high density plasma source, we developed a Silane/Oxygen PECVD process focusing on step coverage deposition rate and uniformity while keeping the breakdown voltage at high values.
Deposition of 2µm thick oxide layers at deposition rate in excess of 500 nm/mn +/- 5% on 8" wafers has been performed (Fig.12).

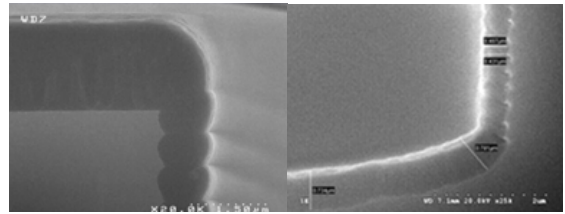

Fig. 12: Top & Bottom corner via with a highly conformal LT-PECVD oxide layer.

An in-depth analysis showed insulating oxide layers with a step coverage higher than 50% for aspect ratio higher than 1:1 with breakdown voltage in excess of 7.5 MV/cm.

## 5. CONCLUSION

A complete range of etching processes has been developed to address not only the "Via First" challenges but also to comply with the Via First before FEOL and Via Last after BEOL requirements. The available process range has been extended to the etching of buried oxide and to the deposition of conformal insulation layers at low temperature. The issues related to the handling of non SEMI Standard wafers has been addressed and successfully implemented onto the latest AMMS products.

## REFERENCES

[1] Emc3d consortium, www.EMC3D.org.

[2] Laermer F., Schilp A., "Method of Anisotropically Etching Si" US patent 5,501,893.

[3] M. Puech, N. Launay, N. Arnal, P. Godinat, JM. Gruffat, "A novel plasma release process and super high aspect ratio process using ICP etching for MEMS", MEMS/NEMS seminar December 3$^{rd}$ 2003.